\RequirePackage{lineno}

\documentclass[aps,prd,onecolumn,superscriptaddress,showpacs,longbibliography]{revtex4-2}
\usepackage{graphicx}
\usepackage{dcolumn}
\usepackage{bm}
\usepackage{hyperref}

\begin{document}

\title{Considerations in the Pursuit of Future Colliders}

\author{Gustaaf Brooijmans}
 \affiliation{Columbia University}

\date{\today}

\begin{abstract}
  Tackling the many open questions in particle physics will require the construction of new colliders.  This short note includes
  a few considerations that seem to be brought up rarely.
\end{abstract}

\maketitle

Particle physics has made enormous strides over the past decades: all particles needed for the Standard Model to be
a complete theory have been observed.  Combined with General Relativity, almost all physical phenomena, from the femtometer
to light-year scales can be explained.  The observed patterns in the Standard Model are not understood, and there is no
confirmed candidate for Dark Matter, something that interacts at least
gravitationally, and has a Compton wavelength at most as large as a typical galaxy.  Similarly, the history of the
universe has unexplained phases, including inflation and the current accelerating expansion.
Basically, from the eV to TeV scale, energies accessible in labs, our knowledge is impressive.  Both below and above that region,
our understanding of our observations is far more limited.

It is conceivable that there are two ``great deserts'': one from about 10$^{-34}$ eV (the size of the universe) to the neutrino mass scale,
and one from about 250 GeV  to the Planck scale, but this seems extremely unlikely.
The reasons for this are many, with some of the
most obvious being the pattern of fermion properties pointing to Grand Unification (e.g.~q$_{\mathrm e}$ = 3q$_{\mathrm down}$
to a very high degree of precision), the strong CP problem likely pointing to another  new symmetry, cosmological evolution indicating
the presence of new fields and/or interactions, neutrino masses hinting at a heavy Majorana scale anywhere between the TeV and Planck scales,
or the Standard Model hierarchy problem which suggests the presence of new physics at a
scale not much higher than the Higgs mass.  A precise study of the Higgs boson itself may yield an understanding of the origin
of the three generations, as this lies in their Yukawa couplings.
It is worth noting that over about 13 orders of magnitude, from the neutrino to the top mass scales,
something interesting happens at typically every order of magnitude, through the presence of particles or characteristic
interaction energy scales\footnote{For example, the 1--100 keV corresponds to atomic energy levels.}.
Exploration of the unknown is a basic requirement towards discovery: to find the next oasis, or the end of the desert,
one has to explore it.

We should thus pursue the most effective ways of understanding the Higgs boson and exploring higher energy scales.  In the immediate future this means exploiting the
(HL-)LHC to the fullest extent possible.  
Whether or
not new physics is seen at the LHC, at least one Higgs factory~\cite{Peskin:2023sed} and one higher energy facility will be needed.  The Snowmass studies pointed to
two options on the distant horizon for the energy frontier: a high-energy muon collider, and a 100 km-scale hadron
collider.  These require more R\&D work, but we do have the knowledge to build and exploit a Higgs factory in the meantime.

None of this is new.  Here are now some less frequently discussed points the community may want to consider in charting a path forward.
\begin{itemize}
\item Ideally we would pursue all options ahead of us.  Muon and hadron collider R\&D will require moderate (but growing) resources, but could we consider
  building both a linear and circular Higgs factory?  One important consideration is the size of the field: how many energy frontier
  physicists are there, and how many does it take?  At the LHC two huge collaborations, two very large collaborations, and a number of smaller
  ones operate $\mathcal{O}$(10) detectors.  This represents $\mathcal{O}$(8000) authors.  A priori electron-positron collider detectors
  are less complex to operate and exploit than e.g.~ATLAS ro CMS, which means the field should be able to build and operate at least three such experiments.
  Others have argued the need for multiple experiments~\cite{Grannis:2023vxf}.
\item What about cost?  So far, the large facilities we have built cost up to about five times the host laboratory's annual operating budget.  The cost estimates shown
  at Snowmass indicate that all of the future colliders under consideration will surpass that number~\cite{Roser:2022sht}.  This likely requires a paradigm shift in our
  approach when bringing this to funding agencies.  We need to recognize that ``for the pursuit of science'' may not be a strong enough argument to
  obtain this level of funding.
  Interactions with congressional aides indicate they
  are far more interested in the benefits to industry (``How many of your students launched a startup?'') than academia (little interest in
  how many became faculty).  Of course, the very definition of fundamental science is that there is no known direct application of our discoveries.
  So, what are the {\em tangible} benefits of building high energy colliders to the economy, or the human community
  as a whole?  A small number of people have made attempts at quantifying this~\cite{Womersley2014,Florio:2016uma,CASTELNOVO20181853}, and their results have always pointed to a
  return on investment larger
  than one.  This number may be going up as the need for data scientists increases.  The field is likely to benefit from a community-supported, and even externally vetted
  presentation of its contributions to society beyond fundamental science.
\item Projects that run late, or indefinitely, inevitably impact the realization of other projects, making planning difficult and unreliable, since
 anticipated resources become unavailable.  To take a concrete example, by 2040, the LHC luminosity
 doubling time will close in on a decade\footnote{There is an exception to this: in a scenario where a signal starts appearing in a final state that is difficult to trigger,
   the experiments may be able to reconfigure their trigger menus to increase that specific dataset much faster.},
 quickly reducing the value of further data taking.  By the start of Run-4, an end game strategy
 should probably be set for LHC operations, based on both expected LHC scientific output and impact on future projects.
\item The environmental impact of large facilities needs to be mitigated.  The world around us is ``retooling'' at an accelerating pace and our field needs to
  find imaginative solutions to operate facilities that need GigaWatts of power (computing included) in a more efficient way.  At Snowmass
  we heard many ideas, for example using the cryogenic infrastructure that is required anyway to store energy when production from renewable sources
  is low.  R\&D on energy recovery linacs is progressing.  Efforts to reduce our energy dissipation need much increased support.
\item We should be aware of the evolved role of conferences in our field, and adjust their organization correspondingly.  There is manifestly no need to go to
  conferences to learn about the latest results.  Instead, conferences now serve three main functions: provide an opportunity for people to showcase
  their work (which is particularly important given our authorship system), provide an educational forum where people can learn about a broader
  set of topics than those they are involved in, and provide time when people are away from their offices and can interact with others in the field to learn
  and generate new ideas.  If this is indeed what conferences are for, then maybe speakers at conferences shouldn't be asked to cover 12 analyses in 15 minutes,
  and conferences should strive to have broad coverage.  
\end{itemize}

\bibliographystyle{LesHouches2}

\bibliography{vision}

\end{document}